\documentclass[twocolumn,amsmath,amssymb]{revtex4}

\usepackage{amsmath}
\usepackage{graphicx}

\begin{document}
\title{Strain gauge fields for rippled graphene membranes under central mechanical load:\\
an approach beyond first-order continuum elasticity}
\author{James V. Sloan,$^1$ Alejandro A. Pacheco Sanjuan,$^{2}$
Zhengfei Wang,$^3$ Cedric Horvath,$^1$ and Salvador Barraza-Lopez$^1$}
\email{sbarraza@uark.edu}
\affiliation{1. Department of Physics. University of Arkansas.
Fayetteville, AR 72701, USA\\
2. Departamento de Ingenier{\'\i}a Mec\'anica. Universidad del Norte. Km.~5 V{\'\i}a Puerto Colombia. Barranquilla, Colombia\\
3. Department of Materials Science and Engineering. University of Utah. Salt Lake City, UT 84112, USA}

\begin{abstract}
We study the electronic properties of rippled freestanding graphene membranes under central load from a sharp tip. To that end, we develop a gauge field theory on a honeycomb lattice valid beyond the continuum theory. Based on the proper phase conjugation of the tight-binding pseudospin Hamiltonian, we develop a method to determine conditions under which continuum elasticity can be used to extract gauge fields from strain. Along the way, we resolve a recent controversy on the theory of strain engineering in graphene: There are no K-point dependent gauge fields. We combine this lattice gauge field theory with atomistic calculations and find that for moderate load, the rippled graphene membranes conform to the extruding tip without significant increase of elastic energy. Mechanical strain is created on a membrane only after a certain amount of load is exerted. In addition, we find that the deformation potential --even when partially screened-- induces qualitative changes on the electronic spectra, with Landau levels giving way to equally-spaced peaks.
\end{abstract}
\date{Published 26 April 2013}
\maketitle

\section{Introduction}
 The interplay of electronic  and mechanical properties of graphene membranes is a subject under intense experimental and theoretical investigation \cite{Nature2007,McEuen1,Pereira1,GuineaNatPhys2010,Kitt2012,Ando2002,vozmediano,deJuanNatPhys,deJuanPRL2012}. Mechanical strain induces gauge fields in graphene that affect the dynamics of charge carriers \cite{Ando2002,Pereira1,GuineaNatPhys2010,vozmediano,Kitt2012}. As graphene can sustain elastic deformations as large as 20\% \cite{nature457_706}, the resulting pseudo-magnetic fields are much larger than those magnetic fields available in state-of-the-art experimental facilities (for example, the highest magnetic field created at the US National High Magnetic Field Laboratory is slightly larger than 100 Tesla). The presence of a  pseudo-magnetic field is observed via broad Landau levels (LLs) in strained graphene nanobubbles on a metal substrate \cite{Crommie}. In addition to the pseudo-magnetic vector potential $\mathbf{A}_s$, strain also induces a scalar deformation potential $E_s$ \cite{Ando2002,deJuanPRB,YWSon} that affects the electron dynamics in complex ways.

 The theoretical formalism has been laid out within the context of first-order continuum elasticity \cite{Ando2002,Pereira1,GuineaNatPhys2010,vozmediano,Kitt2012,deJuanPRB,deJuanNatPhys,deJuanPRL2012}. It is possible to improve the theory from a mechanical perspective. Such a development, provided on the present manuscript, improves our physical understanding of the inter-relation between mechanics and the electrons in graphene. The purpose of the present paper is twofold: First we motivate, build, and validate a novel framework to lay out a theory valid beyond first-order continuum mechanics. This novel formulation brings to the spotlight some of the inherent assumptions of the prevailing theoretical framework; assumptions that have remained to some extent hidden within the continuum formalism. We disclose upfront that the formalism does not take into account the effects of curvature within the framework of Refs.~\onlinecite{vozmediano,deJuanPRB,deJuanNatPhys,deJuanPRL2012}; we will address such shortcoming in the near future. Nevertheless, the reader will realize that the inherent formulation of the theory on the present paper remains novel, bringing a deeper understanding of the formalism for studying the effects of mechanical strain on the electronic properties of graphene.

 Following recent experimental developments in which graphene membranes are studied with local scanning tunneling microscopy probes \cite{us,stmNanoscale2012,stroscio}, our second goal is to demonstrate this novel formalism on freestanding graphene membranes under load by a sharp tip. The input for this formalism is direct `raw' atomic displacements upon strain, as opposed to the always present continuum deformation field $\mathbf{u}(x,y)$ \cite{Nature2007,McEuen1,Pereira1,GuineaNatPhys2010,Kitt2012,vozmediano,deJuanPRL2012}.

 The presentation is given in modular and self-contained form. Hence, discussion of the formalism is given first, then the pure mechanics of freestanding membranes is presented, and predictions from the formalism as pertains to freestanding membranes follows. This helps in focusing either on the basic formulation, or on the predictions from this theory on a experimentally-relevant system. Conclusions are given at the end of the manuscript.

 \section{What are the underlying assumptions of the theory?}
In order to motivate the developments presented here, we express in an explicit form the underlying assumptions of the theory, which can be found as opening statements in Ref.~\onlinecite{GuineaNatPhys2010}: ``{\em If} a mechanical strain varies smoothly on the scale of interatomic distances, it does not break sublattice symmetry but rather deforms the Brillouin zone in such a way that the Dirac cones located in graphene at points $K$ and $K'$ are shifted in opposite directions.'' (See also Ref.~\onlinecite{castroRMP}.)

Previous statement tells us that --provided strain preserves sublattice symmetry-- one can understand the effects of mechanical strain on the electronic structure in terms of a semiclassical approach, as follows: The local strain-induced fields $B_s(\mathbf{r})=\nabla \times A_s(\mathbf{r})$ and $E_s(\mathbf{r})$ are incorporated into a spatially-varying pseudospin Hamiltonian $\mathcal{H}_{ps}(\mathbf{q},\mathbf{r})$, where $\mathcal{H}_{ps}(\mathbf{q})$ is the low-energy expansion of the Hamiltonian in reciprocal space in the absence of strain. We will mention a number of times that the semiclassical approximation is justified {\em if} the strain is slowly varying, that is, when it extends over many unit cells \cite{Ando2002} and preserves sublattice symmetry \cite{GuineaNatPhys2010,vozmediano}.

 Evidently, it is also possible to determine the electronic properties directly from a tight-binding Hamiltonian $\mathcal{H}$ in real space, without resorting to the semiclassical approximation and without imposing an a priori lattice symmetry. That is, while the semiclassical $\mathcal{H}_{ps}(\mathbf{q},\mathbf{r})$ is defined in reciprocal space (thus assuming some reasonable preservation of crystalline order), the tight-binding Hamiltonian $\mathcal{H}$ in real space is more general and can be used for membranes with arbitrary spatial distribution and magnitude of the strain.

  Constituting one of the main arguments of the present paper, we show how to determine if mechanical distortions preserve the fundamental sublattice symmetry. We do this by computing, at each unit cell, angular $\Delta \alpha$ and length changes $\Delta L$ from the adequate nearest-neighbor vectors. Such measures will become relevant for the strongly inhomogeneous strain created by local probes \cite{us,stmNanoscale2012,stroscio}, and will give a quantitative meaning --for the first time-- to statements such as ``long-range mechanical distortion\cite{Ando2002}'' and ``mechanical distortions preserving sublattice symmetry \cite{GuineaNatPhys2010}.'' The program comes down to re-expressing the theory beyond continuum elasticity and explicitly on the atomic lattice, such that matters of spatial scale can be analyzed. For clarity we say that --as a matter of definition-- there is no explicit information of interatomic distances on a continuum media, so sublattice symmetry cannot be determined on this formulation of the theory.

 Indeed, in the only known formulation of the theory (commonly referred to as the {\em lattice}, or {\em tight-binding} approach), both $\mathbf{A}_s$ and $E_s$ are expressed in terms of a {\em continuous} displacement field $\mathbf{u}(x,y)$ obtained within first-order continuum elasticity (CE) \cite{Ando2002,Pereira1,castroRMP,GuineaNatPhys2010,vozmediano}. It is not possible to assess sublattice symmetry on a continuum media, and therefore proper phase conjugation of pseudospin Hamiltonians becomes an implicit assumption of the theory. Continuum elasticity is based on the fundamental assumption, known as Cauchy-Born rule (CBR), that deformations around any material point are homogeneous. But CBR does not hold exactly on the honeycomb lattice, nor under central load or rippling \cite{Ericksen}.

 As an additional contribution on the present paper that adds physical value to our formulation, we mention that an argument was made in the recent past for the inclusion of additional K-point dependent terms to pseudo-magnetic fields \cite{Kitt2012}. Working directly on the atomic lattice, it is easy to show that such terms vanish to first order.
We will also show how the formalism based on CE \cite{Ando2002,Pereira1,castroRMP,GuineaNatPhys2010,vozmediano} becomes a limiting case of the one presented here, when the distortion at all unit cells is small in comparison to the lattice constant $a_0$.

 \section{Formulating a theory beyond first-order continuum elasticity}
\subsection{Sublattice symmetry and measures for long-range mechanical strain}

\begin{figure}[bt]
\includegraphics[width=0.49\textwidth]{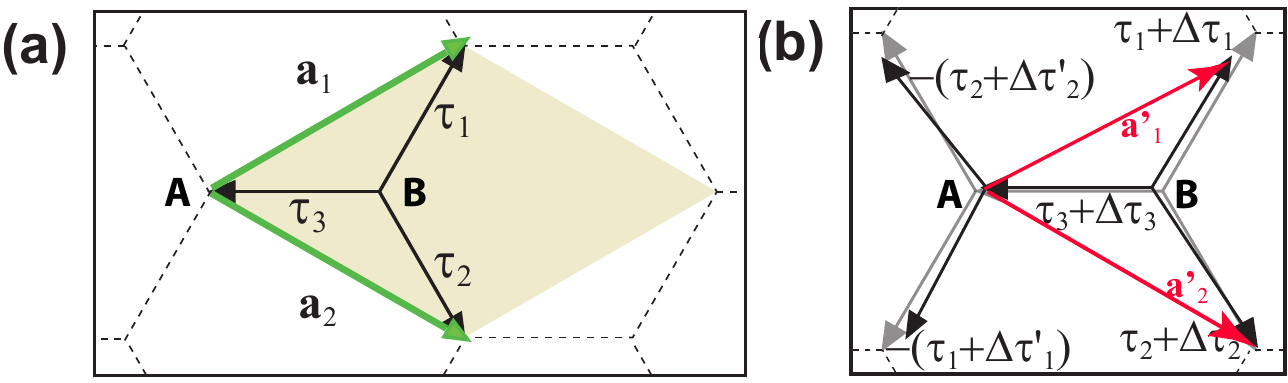}
 \caption{Color online. (a) Unit cell (shaded); lattice vectors $\mathbf{a}_1$ and $\mathbf{a}_2$; and nearest-neighbor vectors $\boldsymbol{\tau}_1$, $\boldsymbol{\tau}_2$, and $\boldsymbol{\tau}_3$. (b) Breakdown of CBR: The nearest-neighbor vectors for the atoms $A$ and $B$ under load are not necessarily mirror-symmetric. If such vectors preserve sublattice symmetry to a reasonable extent, then the lattice vectors can be renormalized univocally to become $\mathbf{a}'_1$ and $\mathbf{a}'_2$, and a theory for strain engineering can be laid out at each and all unit cells.}\label{fig:F1v2}
\end{figure}

Consider the tight-binding Hamiltonian in reciprocal space with no strain \cite{castroRMP}:
\begin{equation}\label{eq:eq01}
\mathcal{H}_0=\left(
\begin{matrix}
0 & -t\sum_{j=1}^3e^{-i\mathbf{k}\cdot\boldsymbol{\tau}_j}\\
-t\sum_{j=1}^3e^{i\mathbf{k}\cdot\boldsymbol{\tau}_j} & 0\\
\end{matrix}
\right),
\end{equation}
 with $t=2.7$ eV. The relevant vectors are shown in Fig.~\ref{fig:F1v2}. Note that in this Figure the zigzag direction lies along the y-axis (a more common choice \cite{GuineaNatPhys2010,vozmediano} is to have the zigzag direction parallel to the x-axis; this is a minor detail, that has to be kept in mind when comparing our final expressions for gauge fields to previous ones \cite{GuineaNatPhys2010,vozmediano}.)
  With the choices for the lattice vectors made in Fig.~\ref{fig:F1v2}(a) we have $\mathbf{b}_1=(1/\sqrt{3},1)2\pi/a_0$, $\mathbf{b}_2=(1/\sqrt{3},-1)2\pi/a_0$. To test the purported K-point dependent correction to the theory \cite{Kitt2012}, we write down all six K-points explicitly:
  \begin{eqnarray}\label{eq:defK}
  \mathbf{K}_1=(\mathbf{b}_1-\mathbf{b}_2)/3=(0,1)4\pi/(3a_0),\\
  \mathbf{K}_2=(2\mathbf{b}_1+\mathbf{b}_2)/3=(\sqrt{3},1)2\pi/(3a_0),\nonumber\\
  \mathbf{K}_3=(\mathbf{b}_1+2\mathbf{b}_2)/3=(\sqrt{3},-1)2\pi/(3a_0). \nonumber
  \end{eqnarray}
  It follows that:
  \begin{equation}\label{eq:defK2}
  \mathbf{K}_{n+3}=-\mathbf{K}_{n}, \text{$ ({n}=1,2,3)$},
  \end{equation}
 and $\mathbf{k}=\mathbf{K}_n+\mathbf{q}$. The low-energy expansion of Eqn.~\eqref{eq:eq01} (when $\mathbf{q}<<\mathbf{K}_n$) expresses the dynamics of pseudospinor on the honeycomb lattice; in that limit $\mathcal{H}_{0}\to \mathcal{H}_{ps}$. Though certainly redundant due to crystal symmetry in the absence of strain, one is free to define one $\mathcal{H}_{ps}$ at each unit cell, with the finite number ($N/2$) of pseudospin Hamiltonians for a membrane with a finite number ($N$) of atoms.

 This finite number of pseudospinor Hamiltonians that can be defined on a membrane with $N$ atoms represents the first departure of our formulation of the theory when compared with the formalism developed on a continuum media:
In the continuum approach \cite{Ando2002,Pereira1,GuineaNatPhys2010,vozmediano,Kitt2012,deJuanPRB,deJuanNatPhys,deJuanPRL2012}, the two degrees of freedom of a given local pseudospin Hamiltonian $\mathcal{H}_{ps}(\mathbf{q},\mathbf{r})$ should correspond to those of an underlying unit cell with two atoms. However, the pseudospin Hamiltonian is defined as a continuous function of coordinates $\mathbf{r}$ and is hence detached from the actual spatial structure of the lattice. This semiclassical approach is justified when the spatial variation of the strain is small on the scale of the lattice constant $a_0$. In the present work, we develop a more general method which preserves the spatial scale of mechanical distortion relative to $a_0$, as well as the total number of local pseudospin Hamiltonians. By doing so we can analyze situations in which the continuum approach breaks down. In addition, we show that pseudo-magnetic vector fields do not depend on $K-$points.

 {\em The only way to know whether the strain preserves sublattice symmetry \cite{GuineaNatPhys2010} is by analyzing relative atomic displacements.} The nearest neighbor vectors for atom $A$ ($B$) become:
 $-\boldsymbol{\tau}_1-\Delta\boldsymbol{\tau}_1'$, $-\boldsymbol{\tau}_2-\Delta\boldsymbol{\tau}_2'$, and $-\boldsymbol{\tau}_3-\Delta \boldsymbol{\tau}'_3$ ($\boldsymbol{\tau}_1+\Delta\boldsymbol{\tau}_1$, $\boldsymbol{\tau}_2+\Delta\boldsymbol{\tau}_2$, and $\boldsymbol{\tau}_3+\Delta\boldsymbol{\tau}_3$);
 see Fig.~\ref{fig:F1v2}(b). While $\Delta \boldsymbol{\tau}_3=\Delta \boldsymbol{\tau}'_3$ by construction, $\Delta\boldsymbol{\tau}_{1(2)}$ is not necessarily equal to $\Delta\boldsymbol{\tau}'_{1(2)}$ for arbitrary strain.
To better quantify the local departures from the sublattice symmetry at any given unit cell, we define the differences in angular orientation $\Delta \alpha_j$ and length $\Delta L_j$ for nearest-neighbor vectors under mechanical load: Writing $\Delta \boldsymbol{\tau}_j=(\Delta x_j,\Delta y_j,\Delta z_j)$ and $\Delta \boldsymbol{\tau}'_j=(\Delta x'_j,\Delta y'_j,\Delta z'_j)$ for  $j=1,2$:
\begin{equation}\label{eq:beta}
\small(\boldsymbol{\tau}_j+\Delta  \boldsymbol{\tau}_j)\cdot(\boldsymbol{\tau}_j+\Delta\boldsymbol{\tau}'_j)=
|\boldsymbol{\tau}_j+\Delta\boldsymbol{\tau}_j||\boldsymbol{\tau}_j+\Delta\boldsymbol{\tau}'_j|\cos(\Delta\alpha_j),
\end{equation}
\begin{equation}\label{eq:sign}
\small\text{sgn}(\Delta \alpha_j)=\text{sgn}\left([(\boldsymbol{\tau}_j+\Delta\boldsymbol{\tau}_j)
\times(\boldsymbol{\tau}_j+\Delta\boldsymbol{\tau}'_j)]\cdot \hat{k}\right),\end{equation}
where $\hat{k}$ is a unit vector along the z-axis, and:
\begin{equation}\label{eq:L}
\small
\Delta L_j\equiv |\boldsymbol{\tau}_j+\Delta\boldsymbol{\tau}_j|-|\boldsymbol{\tau}_j+\Delta\boldsymbol{\tau}'_j|.
\end{equation}
We reiterate that the existing theory\cite{Ando2002,GuineaNatPhys2010,vozmediano} {\em requires} sublattice symmetry to hold: $\Delta\alpha_j\simeq 0$, and $\Delta L_j\simeq 0$. In practice however, as no measure existed to test those requirements, in applying the theory one actually is lead to {\em assume a priori} that $\Delta\alpha_j=0$, and $\Delta L_j= 0$.

 Later on, we will have the opportunity to see how
 quantifiable deviations of sublattice symmetry occur under central load. Spatial locations where $\Delta \alpha_j$ and $\Delta L_j$ are much larger than zero indicate that the continuum theory ceases to be applicable there, as the lack of sublattice symmetry will not allow proper phase conjugation of pseudospin Hamiltonians. This should not be even surprising because for a reciprocal space to exist one has to preserve the crystal symmetry. When the crystal symmetry is strongly perturbed, the reciprocal space representation looses its physical meaning. In such scenario $\mathcal{H}$ (and hence the LDOS) is still meaningful, and so is $E_s$, but $\mathbf{A}_s$ starts to be ill-defined as a local lack of sublattice symmetry necessarily implies the lack of proper phase conjugation.
 Unfortunately, it is not always possible to express distortions using a continuum theory for lattices with inner structure (sublattices $A$ and $B$): {\em first-order CE breaks down} \cite{Ericksen}, and inhomogeneity of the atomic displacements --reflecting lack of periodicity upon non-uniform strain and shown in Fig.~\ref{fig:F1v2}-- sets in.

\subsection{Relative shift of the $K$ and $K'$ points upon strain}

 In the more general and lattice-explicit approach being presented here, $\mathbf{A}_s$ can be obtained at unit cells in which $\Delta\alpha_j\simeq 0$ and $\Delta L_j\simeq 0$,  by a (local) replacement of $\boldsymbol{\tau}_j$ in Eqn.~\eqref{eq:eq01} with displaced vectors at each of the two sublattice atoms. One realizes that under load the lattice vectors become $\mathbf{a}_1'=\tau_1+\Delta \tau_1 -\tau_3-\Delta \tau_3$; $\mathbf{a}_2'=\tau_2+\Delta \tau_2 -\tau_3-\Delta \tau_3$, which in turn leads to renormalized $\mathbf{K}_n$ points. So, to first order in displacements, the reciprocal lattice vectors can be obtained from:
 \begin{equation}
 \mathcal{B}'\simeq 2\pi\left(\mathcal{A}^{-1}-\mathcal{A}^{-1}\Delta \mathcal{A}\mathcal{A}^{-1}\right)^T,
 \end{equation}
 where $\mathcal{A}=(\mathbf{a}_1^T, \mathbf{a}_2^T)$, and $\Delta\mathcal{A}=([\mathbf{a}_1'-\mathbf{a}_1]^T, [\mathbf{a}_2'-\mathbf{a}_2]^T)$. The form of $\mathcal{B}'$ is convenient, as then $\mathbf{b}_{1,2}'=\mathbf{b}_{1,2}+\Delta \mathbf{b}_{1,2}$ and therefore using Eqn.~\eqref{eq:defK} one gets:
 \begin{equation}\label{eq:deltaK}
 \mathbf{K}_n'=\mathbf{K}_n'(\{\Delta \boldsymbol{\tau}_j\})=\mathbf{K}_n+\Delta \mathbf{K}_n(\{\Delta \boldsymbol{\tau}_j\})
 \end{equation}
 as well.

  Now, there are three choices for defining a set of $K$ and $K'$ pairs: $K \equiv \mathbf{K}_n$ and $K'\equiv \mathbf{K}_{n+3}$ with $n=1,3$. So Equation \eqref{eq:deltaK}, in combination with Eqn.~\eqref{eq:defK2} tells us that:
  \begin{equation}
  \Delta \mathbf{K}_n(\{\Delta \boldsymbol{\tau}_j\})=-\Delta \mathbf{K}_{n+3}(\{\Delta \boldsymbol{\tau}_j\}),
  \end{equation}
   so that ``the Dirac cones located in graphene at points $K$ and $K'$ are shifted in opposite directions \cite{GuineaNatPhys2010,castroRMP}.'' This fact builds onto the consistency of the present formulation of the theory.

\subsection{Lattice-explicit strain-gauge potentials (negligible curvature)}
 How does the pseudospinor Hamiltonians look in the new formalism? Standard manipulation ( $\boldsymbol{\tau}_j \to \boldsymbol{\tau}_j+\Delta\boldsymbol{\tau}_j$; $\mathbf{k}=\mathbf{K}_n'+\mathbf{q}$; $j=1,2,3$, $n=1,...,6$) leads to the off-diagonal term:
\begin{eqnarray}\label{eq:eq03}\small
\sum_{j=1}^3-(t+\delta t_{j})e^{i(\mathbf{K}_n+\Delta \mathbf{K}_n +\mathbf{q})\cdot(\boldsymbol{\tau}_j+\Delta \boldsymbol{\tau}_j)}\simeq\nonumber\\
\small\sum_{j=1}^3-(t+\delta t_{j})
e^{i\mathbf{K}_n\cdot\boldsymbol{\tau}_j}
e^{i\Delta\mathbf{K}_n\cdot\boldsymbol{\tau}_j}
e^{i\mathbf{K}_n\cdot\Delta\boldsymbol{\tau}_j}
e^{i\mathbf{q}\cdot\boldsymbol{\tau}_j},
\end{eqnarray}
Here, $\delta t_j$ is the change of the hopping parameter upon strain. The other off-diagonal term is:
\begin{equation}\label{eq:eq032}\small
\sum_{j=1}^3-(t+\delta t_{j}')
e^{-i\mathbf{K}_n\cdot\boldsymbol{\tau}_j}
e^{-i\Delta\mathbf{K}_n\cdot\boldsymbol{\tau}_j}
e^{-i\mathbf{K}_n\cdot\Delta\boldsymbol{\tau}'_j}
e^{-i\mathbf{q}\cdot\boldsymbol{\tau}_j},
\end{equation}
(note that in Eqn.~\ref{eq:eq032}, $\Delta\mathbf{K}_n$ is expressed in terms of unprimed $\Delta \boldsymbol{\tau}_j$'s).
 When $\Delta\boldsymbol{\tau}_{1(2)}\ne\Delta\boldsymbol{\tau}_{1(2)}'$ it follows that $\delta t_{1(2)}\ne \delta t_{1(2)}'$. If the sublattice symmetry does not hold to measurable extent, Eqn.~\eqref{eq:eq03} would not be exactly conjugated to Eqn \eqref{eq:eq032}, and applicability of the theory at those unit cells is questionable.

Only if $\Delta\alpha_j\simeq 0$ and $\Delta L_j\simeq 0$, discrete versions of $\mathcal{H}_{ps}$ (and hence $\mathbf{A}_s$) can be extracted from lowest-order expansions of Eqn.~\eqref{eq:eq03} and \eqref{eq:eq032} with $\Delta\boldsymbol{\tau}_{1(2)}'$ replaced by $\Delta\boldsymbol{\tau}_{1(2)}$:
\begin{equation}\small
\sum_{j=1}^3-(t+\delta t_j)e^{i\mathbf{K}_n\cdot\boldsymbol{\tau}_j}[1+i(
\Delta \mathbf{K}_n\cdot\boldsymbol{\tau}_j+\mathbf{K}_n\cdot\Delta\boldsymbol{\tau}_j+\mathbf{q}\cdot\boldsymbol{\tau}_j)].
\end{equation}
Remarkably, the term:
\begin{eqnarray}\label{eq:linear}\small
&\sum_{j=1}^3-te^{i\mathbf{K}_n\cdot\boldsymbol{\tau}_j}[1+i(
\Delta \mathbf{K}_n\cdot\boldsymbol{\tau}_j+\mathbf{K}_n\cdot\Delta\boldsymbol{\tau}_j+\mathbf{q}\cdot\boldsymbol{\tau}_j)]\nonumber\\
&=\sum_{j=1}^3-te^{i\mathbf{K}_n\cdot \boldsymbol{\tau}_j}(1+i\mathbf{q}\cdot\boldsymbol{\tau}_j),
\end{eqnarray}
 leads to the linear dispersion because the phasors on $\sum_{j=1}^3e^{i\mathbf{K}_n\cdot\boldsymbol{\tau}_j}(
\Delta \mathbf{K}_n\cdot\boldsymbol{\tau}_j+\mathbf{K}_n\cdot\Delta\boldsymbol{\tau}_j)$ add up to zero (this can be shown by explicit calculation). Neglect of the term linear on $\Delta \mathbf{K}_n$ in Ref.~\cite{Kitt2012} led to artificial gauges. Hence, it is just the single term:
\begin{equation}\label{eq:gauge}\small
\sum_{j=1}^3-\delta t_j e^{i\mathbf{K}_n\cdot \boldsymbol{\tau}_j},
\end{equation}
 that leads to pseudo-magnetic gauge field to lowest-order.

 When the zigzag direction is parallel to the $x-$axis \cite{GuineaNatPhys2010,vozmediano}, the real (imaginary) part of Eqn.~\ref{eq:gauge} directly leads to the x- (y-)component of $\mathbf{A}_s$. With the choice made in Fig.~\ref{fig:F1v2}(a)
  one obtains reverted components:
\begin{equation}\label{eq:As}
\small\mathbf{A}_s=\frac{\phi_0}{\pi a_0}
\left(
\begin{smallmatrix}
\frac{\delta t_1-\delta t_2}{t}\\
\pm\frac{-\delta t_1-\delta t_2+2\delta t_3}{\sqrt{3}t}
\end{smallmatrix}
\right),
\end{equation}
with $\phi_0=h/2e$ the flux quantum. The `+' sign appears for $\mathbf{K}_1$, $\mathbf{K}_3$ and $\mathbf{K}_5$; the `$-$' sign (implying $y\to-y$) near $\mathbf{K}_2$, $\mathbf{K}_4$ and $\mathbf{K}_6$. The net pseudo-magnetic field is zero. Dirac's eqn. in terms of $\mathcal{H}_{ps}$ is ($\hbar v_F\equiv t\sqrt{3}a_0/2$):
\begin{equation}\label{eq:new4}
\small \mathcal{H}_{ps}\Psi=\hbar v_F \boldsymbol{\sigma} \cdot \left(\mathbf{q}-\frac{e\mathbf{A}_s}{\hbar}\right)\Psi+\mathbf{I}E_s\Psi,
\end{equation}
with $\mathbf{I}$ the $2\times 2$ identity, and $\delta t_j=-|\beta|t\boldsymbol{\tau}_j\cdot\Delta \boldsymbol{\tau}_j/a_0^2$
(Eqn.~3.7 in Ref.~\cite{Ando2002}, or Eqn.~56 in Ref.~\cite{vozmediano}). $|\beta|=-\frac{\partial \ln t}{\partial \ln a_0}\simeq 2.3$ \cite{Ando2002,GuineaNatPhys2010,vozmediano}, hence we arrive at:
\begin{equation}\label{eq:deltat}
\small\mathbf{A}_s=
\frac{-\phi_0|\beta|}{\pi a_0^3}
\left(
\begin{matrix}
\boldsymbol{\tau}_1\cdot\Delta \boldsymbol{\tau}_1-\boldsymbol{\tau}_2\cdot\Delta \boldsymbol{\tau}_2\\
\pm\frac{-\boldsymbol{\tau}_1\cdot\Delta \boldsymbol{\tau}_1-\boldsymbol{\tau}_2\cdot\Delta \boldsymbol{\tau}_2
+2\boldsymbol{\tau}_3\cdot\Delta \boldsymbol{\tau}_3}{\sqrt{3}}
\end{matrix}
\right),
\end{equation}
so at a given unit cell, each component of $\mathbf{A}_s$ takes {\em a single value}.
 We assume $E_s$ to be linearly-dependent to the average bond increase \cite{YWSon}:
\begin{equation}\label{eq:ED}\small
  E_s(\mathbf{r})=-\frac{0.3 \text{ }eV}{0.12}\frac{1}{3}\sum_{j=1}^3\frac{|\boldsymbol{\tau}_j+\boldsymbol{\Delta \tau}_j|-a_0/\sqrt{3}}{a_0/\sqrt{3}}.
\end{equation}
[$E_s$ in Fig.~\ref{fig:F4}(a) is similar to the profile in Ref.~\cite{deJuanPRB}.] Eqns.~\eqref{eq:deltat} and \eqref{eq:ED} express the gauge fields in terms of lattice displacements, representing one of our main results. Besides the inherent physical motivation which has been explained in detail, Eqns.~\eqref{eq:deltat} and \eqref{eq:ED} obviate the need for a continuous deformation field, and hold regardless of the magnitude of the deformation, even in the anharmonic regime (refer to Fig.~\ref{fig:F2}(a)).

\subsection{Limiting form of the vector potential for strain varying slowly with respect to $a_0$}

In the limit $|\Delta \boldsymbol{\tau}_j|/a_0\to 0$ the theory from CE is restored. Indeed,
\begin{equation}
\Delta \boldsymbol{\tau}_j^T=(\Delta x_j, \Delta y_j)^T\to
\left(
\begin{smallmatrix}
u_{xx}& u_{xy}\\
u_{xy}& u_{yy}
\end{smallmatrix}
\right)\boldsymbol{\tau}^T_j \text{ (Cauchy-Born rule)},
\end{equation}
 and after simple algebraic manipulations one gets:
\begin{equation}\label{eq:gcont}
\mathbf{A}_s\to\frac{|\beta|\phi_0}{2\pi\sqrt{3}a_0}
\left(
\begin{smallmatrix}
-u_{xy}\\
\pm\frac{u_{yy}-u_{xx}}{2}
\end{smallmatrix}
\right).
\end{equation}

\begin{figure}[bt]
\includegraphics[width=0.49\textwidth]{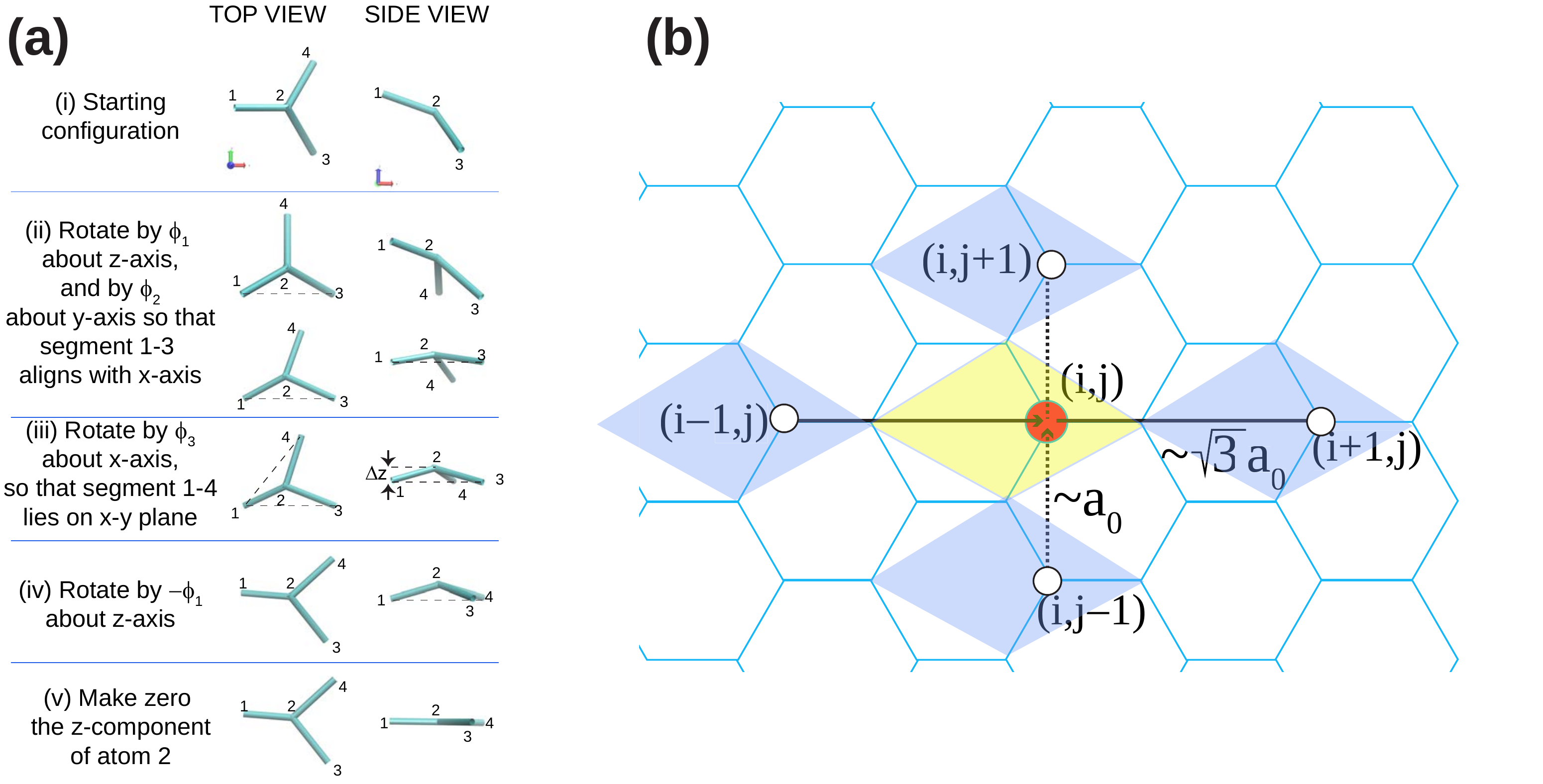}
 \caption{Color online. (a) The process to produce two-dimensional displacements from three-dimensional ones. (b) Locations from which finite differences are computed.}\label{fig:flattening}
\end{figure}

The novel formalism has been completely motivated, laid out, and validated at this moment.
Now, to plot $\mathbf{B}_s$ a flattening procedure and a method of finite differences were developed so that the three-dimensional $\Delta \boldsymbol{\tau}$'s from atomic displacements could be used in Eqns.~(\ref{eq:new4}) and (\ref{eq:deltat}). See Figure \ref{fig:flattening}(b).

 The idea is to flatten locally the three nearest neighbor vectors from their positions under stress such that their lengths and relative angles are preserved to the greatest extent possible. Flattened vectors are then used in the two-dimensional $\mathcal{H}_{ps}$. It should be clear that this procedure can only be appropriate when the vectors $\Delta\boldsymbol{\tau}_j$ ($j=1,2,3$) are small with respect to the lattice constant. Extraction of $B_s$ is only sensible at locations away from the tip, where $\Delta\boldsymbol{\tau}_j\simeq\Delta\boldsymbol{\tau}'_j$. The process involves rotating the three vectors first so that the three outer points defining these vectors lie on the x-y plane. This is described as step (i-iii) in Figure 2. Once these points lie on the same plane, we perform an additional rotation about the z-axis so that one of the vectors is close to its original projection along the x-y plane (Figure 2(iv)). The process is completed by bringing the atom in the center towards the x-y plane, by setting its magnitude along the z-axis to be zero (Figure 2(v)).
 The vertical displacement in Fig.~2(iv) is exaggerated (more below).

As said before, $\mathbf{B}_s$ is obtained in terms of finite differences on the flattened membrane, as follows:
\begin{equation}
\mathbf{B}_s =\hat{k}(\Delta_x A_y-\Delta_y A_x),
\end{equation}
with $\hat{k}$ a unit vector pointing out of plane. $A_x$ and $A_y$ are to be computed at {\em unit cells} for which sublattice symmetry is reasonably preserved. The partial derivative is estimated after flattening as:
\begin{eqnarray*}
\Delta_x A_y\simeq \\
\frac{1}{2}
\begin{small}
\left(
\frac{A_y(\mathbf{r}_{i+1,j})-A_y(\mathbf{r}_{i,j})}{|\mathbf{r}_{i+1,j}-\mathbf{r}_{i,j}|}+
\frac{A_y(\mathbf{r}_{i,j})-A_y(\mathbf{r}_{i-1,j})}{|\mathbf{r}_{i,j}-\mathbf{r}_{i-1,j}|}
\right),
\end{small}
\end{eqnarray*}
 and:
\begin{eqnarray*}
\Delta_y A_x\simeq\\
\frac{1}{2}
\begin{small}
\left(
\frac{A_x(\mathbf{r}_{i,j+1})-A_x(\mathbf{r}_{i,j})}{|\mathbf{r}_{i,j+1}-\mathbf{r}_{i,j}|}+
\frac{A_x(\mathbf{r}_{i,j})-A_x(\mathbf{r}_{i,j-1})}{|\mathbf{r}_{i,j}-\mathbf{r}_{i,j-1}|}
\right).
\end{small}
\end{eqnarray*}
See Figure 2(b) for a schematic illustration of the locations involved.
The computation of the curl in terms of finite differences reflects the inherently {\em discrete} nature of the present formulation.
 No other result, including the LDOS, required flattening. We concur that the present method for obtaining $B_s$ could benefit from the ideas within the geometrical approach \cite{vozmediano,deJuanNatPhys,deJuanPRL2012}. We expect to address this aspect in the near future. It should be clear, nevertheless, that our approach beyond continuum elasticity never looses its novelty and value.

Motivated by recent experiments \cite{us,stmNanoscale2012,stroscio}, we illustrate previous considerations by studying the freestanding graphene membranes under central load by a sharp Scanning Tunneling Microscope (STM) tip. As we will show, the membranes are rippled before load because of dynamic (temperature-induced) structural distortions \cite{Fasolino1}, and because of static structural distortions created by interaction with a substrate, the deposition process \cite{Nature2007}, or line stress at edges.

\section{The mechanical behavior of Freestanding graphene membranes under central mechanical load}

 \begin{figure}[h!]
\includegraphics[width=0.49\textwidth]{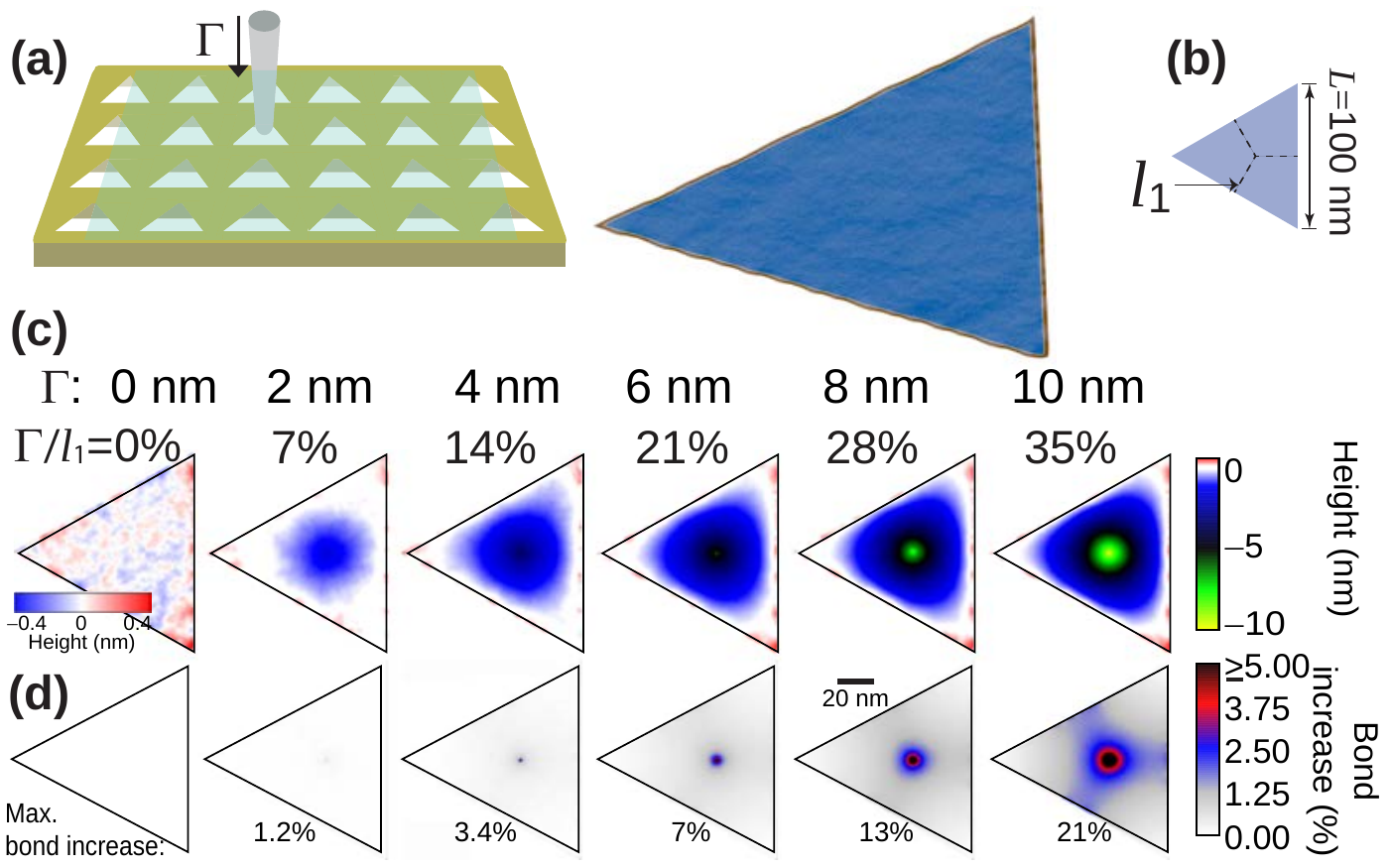}
 \caption{Color online. (a) Triangular graphene membranes under load. (b) $l_1$ is the shortest distance from center to edge. (c) Height profiles and (d) increase of the bond lengths.}\label{fig:F1}
 \end{figure}

\subsection{Details of the systems studied and the molecular dynamics calculations}

 We considered triangular membranes with side $L=100$ nm and 0.16 million atoms [Fig.~\ref{fig:F1}(a)].
We have chosen triangular boundaries since they are known to create the most uniform pseudo-magnetic field \cite{GuineaNatPhys2010}. Equilibrium atomic configurations were obtained from classical molecular dynamics simulations at 1 Kelvin \cite{LAMMPS}. Prior to load, the initially flat membrane is allowed to relieve line strain at its edges, equilibrating forces for 500,000 fs with all atoms moving freely. At this low temperature the average lattice constant is $a_0=2.41$ \AA.
In the initial state after relaxation, the membrane is rippled, with a minimum-to-maximum vertical displacement of 0.8 nm (see leftmost subplot in Fig.~\ref{fig:F1}(c)).

The membrane rims (shown in brown in Fig.~\ref{fig:F1}(a)) represent the mechanical support of a freestanding membrane; they are clamped after the equilibrium rippled conformation is obtained. The height fluctuations seen on the first subplot in Fig.~\ref{fig:F1}(c) tell us that a finite-size graphene membrane behaves as a {\em shell} in equilibrium, because it has nonzero local curvature in the absence of applied strain. This behavior will be necessarily linked to the magnitude of the mechanical strain upon load. The (static) rippling discussed here and due to finite size is different from the dynamic effect produced by temperature \cite{Fasolino1}. We must note that most theoretical works consider as their starting point a planar membrane (a {\em thin plate} in mechanical jargon). Exceptions are presented in the {\em geometrical} approach (Refs.~\cite{deJuanPRB,vozmediano,deJuanPRL2012}), a formulation of the theory still on a continuum media, where higher-order terms --related to curvature-- enter in. Being a theory on a continuum as well, the issue of the scale of the mechanical distortion here discussed carries on.

Strain is induced on the rippled membrane by pushing down a spherical tip (3 nm in diameter), interacting with the membrane via a van der Waals term (details can be provided upon request). The tip pushes the membrane at speed $v=10^{-5}$ nm/fs to a distance  $\Gamma=vT$, where $T$ is the load time.
The load protocol used here is different than the one used in experiments \cite{us,stroscio}, where the tip retracts away from the membrane. The membrane --initially 0.2 nm below the indenter-- deforms as soon as the tip moves down (though the deformation initially preserves interatomic bond distances, more below). After load, the membranes are equilibrated at 1 Kelvin for 500,000 fs, with the tip remaining at a vertical distance $\Gamma$.

\subsection{Membrane mechanics beyond first-order continuum elasticity: The isometric and anharmonic load regimes}

The dimensionless quantity $\Gamma/l_1$ --with $l_1=28.9$ nm the closest distance from the geometrical center to the edge [Fig.~\ref{fig:F1}(b)]-- has been used as a measure of strain \cite{us,stroscio}. It proves inaccurate for freestanding (rippled) membranes as the initial deformation is isometric (i.e., bond changes are initially unnoticeable). To see this, we show in Fig.~\ref{fig:F1}(c) the height profiles versus $\Gamma/l_1$, and in Fig.~\ref{fig:F1}(d) the corresponding increase of the bond lengths. Even though the height plots show some amount of curvature, no significant bond length increase can be seen on the first two plots in Fig.~\ref{fig:F1}(d). Indeed, when $\Gamma/l_1$ is 7\% already, the largest bond increase, right below the tip, is equal to 1.2\% (so that for this amount of load, the bond increase is not equal to $\Gamma/l_1$, but rather to $\sim\Gamma/6l_1$). Thus, neglect of rippling \cite{Nature2007,Fasolino1} on thin-plate-based strain engineering (i.e., setting the initial configuration to be a plate) may lead to overestimating $B_s$, an observation relevant to experimentalists generating strain on freestanding graphene with local probes. The largest bond length increase approaches $\Gamma/l_1$ for higher load, as the distortion below the tip becomes highly nonlinear (more below). Figure \ref{fig:F1}(d) also indicates bond length increases with radial symmetry near the geometrical center, determining the spatial profile of the pseudo-magnetic field that is generated by a spherical tip.

\begin{figure}[tb]
\includegraphics[width=0.49\textwidth]{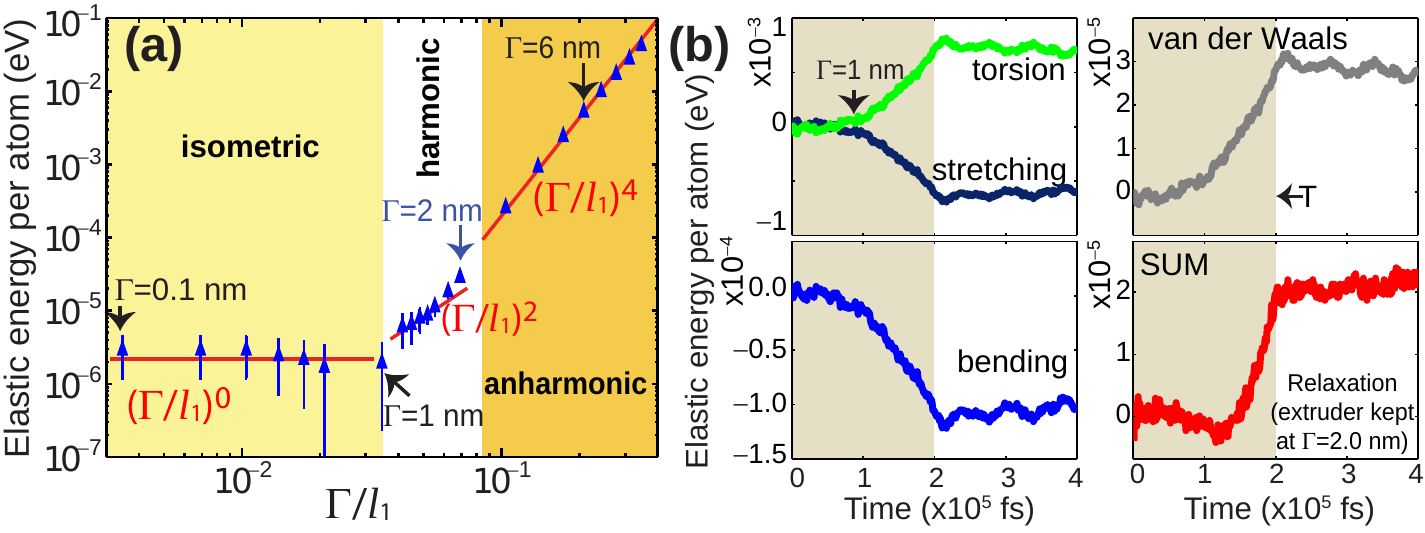}
 \caption{Color online. (a) The elastic energy {\em vs.} indentation shows three distinct regimes: (i) isometric, due to the initially rippled conformation, (ii) linear (or harmonic) and (iii) nonlinear (anharmonic). (b) Decomposition of the elastic energy for a load $\Gamma=2.0$ nm (shaded area represents the load time).
}\label{fig:F2}
\end{figure}

 We perform an analysis of the elastic energy as a function of $\Gamma/l_1$ (Fig.~\ref{fig:F2}(a)). We observe three distinct regimes. In the first regime, the elastic energy does not increase beyond fluctuations signified by error bars: This is the isometric regime, in which the initially rippled membrane follows the probe without necessarily increasing its elastic energy, nor producing significant mechanical strain. This regime holds for values of $\Gamma/l_1$ up to a few percent.
 The second regime in Fig.~\ref{fig:F2}(a) is harmonic, as indicated by a quadratic dependence of elastic energy on $\Gamma/l_1$. The harmonic regime holds for $\Gamma/l_1$ in a narrow range between 4 and 10\%. For $\Gamma/l_1 >$ 10\%, the system enters the anharmonic regime. We note that in the context of thin plates, only the harmonic and anharmonic regimes have been discussed in the past \cite{WHDuan}. The theory based on first-order continuum elasticity may not hold in the anharmonic regime.

The results shown in Fig.~\ref{fig:F2}(a) can be understood by an analysis of the total and constituent elastic energies. The decomposition of the total elastic energy into torsional, stretching, and bending components is shown in Fig.~\ref{fig:F2}(b) for triangular membrane subject to the load $\Gamma=2.0$ nm. The shaded area indicates the load time $T$; atomic relaxation follows in the remaining time. We observe that the total energy does not increase until $\Gamma=1$ nm, however the energy components provide a very interesting insight: The two leading energy contributions are the stretching and torsion of bonds. While the stretching contribution decreases --perhaps due to the fact that the tip suppresses some fluctuations in bond distances when pushing the membrane, we find that the torsion energy goes up by an almost equal amount.
The remaining bending contribution to the elastic energy is an order of magnitude smaller. Thus, the total elastic energy remains practically constant for loads up to $\Gamma=1$ nm.

\section{Applying the formalism to graphene membranes under central load}
\subsection{Evaluation of sublattice symmetry}

 We plot the measures given by Equations (4-6) in Figure 5, in order to demonstrate their actual value. unit cells for which sublattice symmetry
 hold to numerical precision are told by the white color. As expected, deviations become larger in the close proximity of the mechanical extruder (located at the membrane's geometrical center), and  for increasing values of $\Gamma/l_1$.

\begin{figure}[tb]
\includegraphics[width=.49\textwidth]{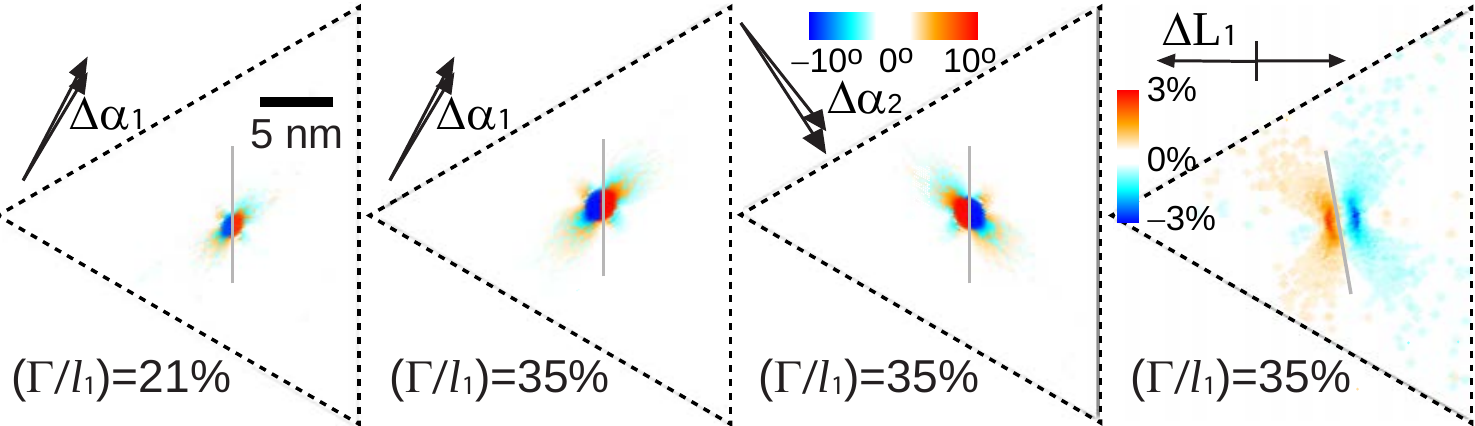}
 \caption{Color online.  Representation of Eqns.~4-6 for our system. Solid lines highlight anti-symmetric patterns.}\label{fig:F5v2}
\end{figure}

\subsection{Evaluation of the flattening procedure}
 The vertical displacement in Fig.~2(d) is exaggerated. This displacement is less than 0.3\% $a_0$ at distances 1 nm away from the extruder, as seen in Fig.~6. This value is one thousand times smaller than $\Gamma/l_1=35$\% employed to generate the atomic configuration, and represents the order of magnitude of the error introduced by the collapsing of the central atom into the x-y plane.
\begin{figure}[tb]
\includegraphics[width=.3\textwidth]{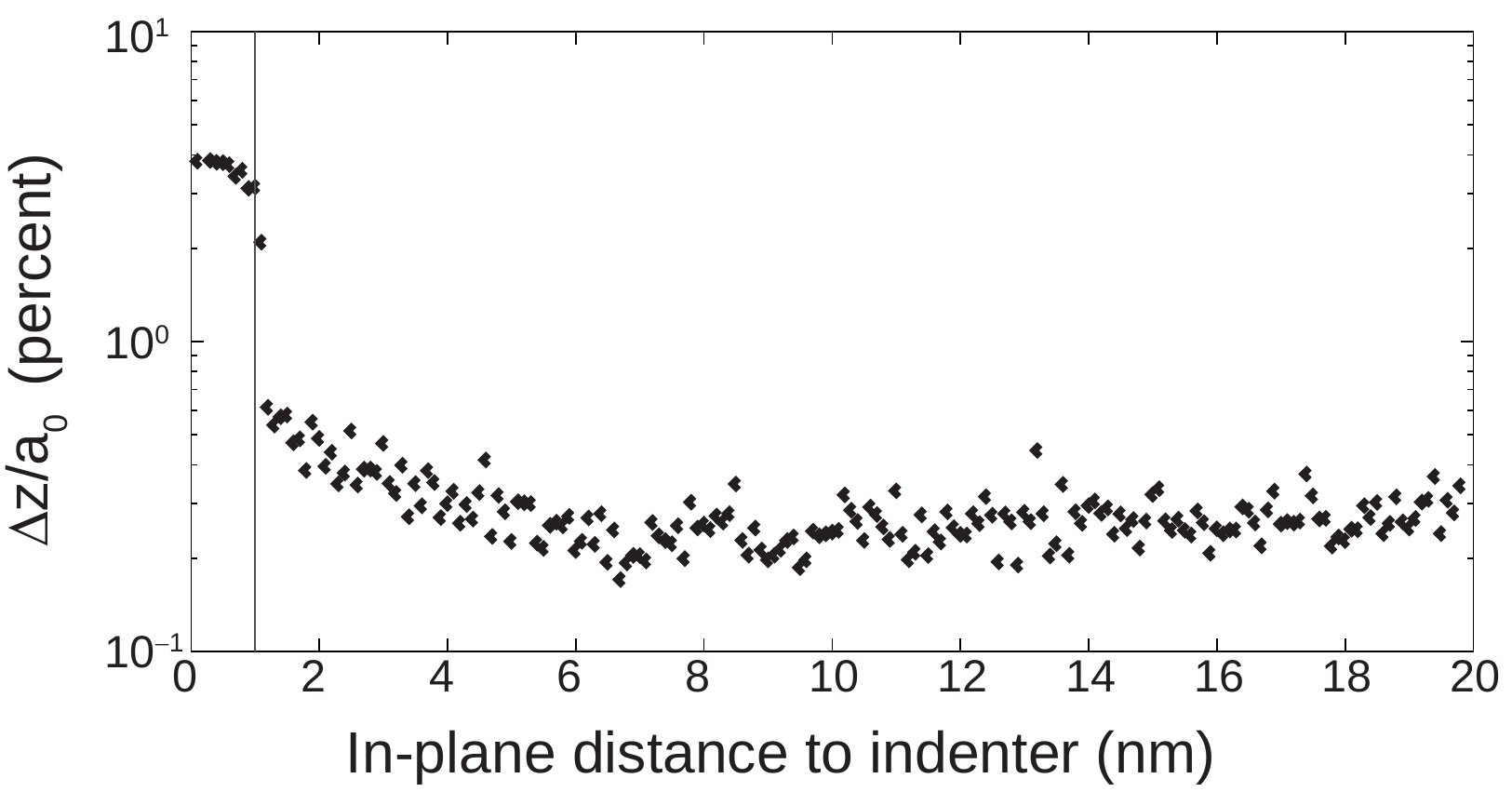}
 \caption{$\Delta z$ versus in-plane distance to extruding tip $d$. $\Delta z\lesssim 0.3$ \% for $d>1$ nm.}\label{fig:F6v2}
\end{figure}

\subsection{Gauge fields}
\begin{figure}[tb]
\includegraphics[width=.49\textwidth]{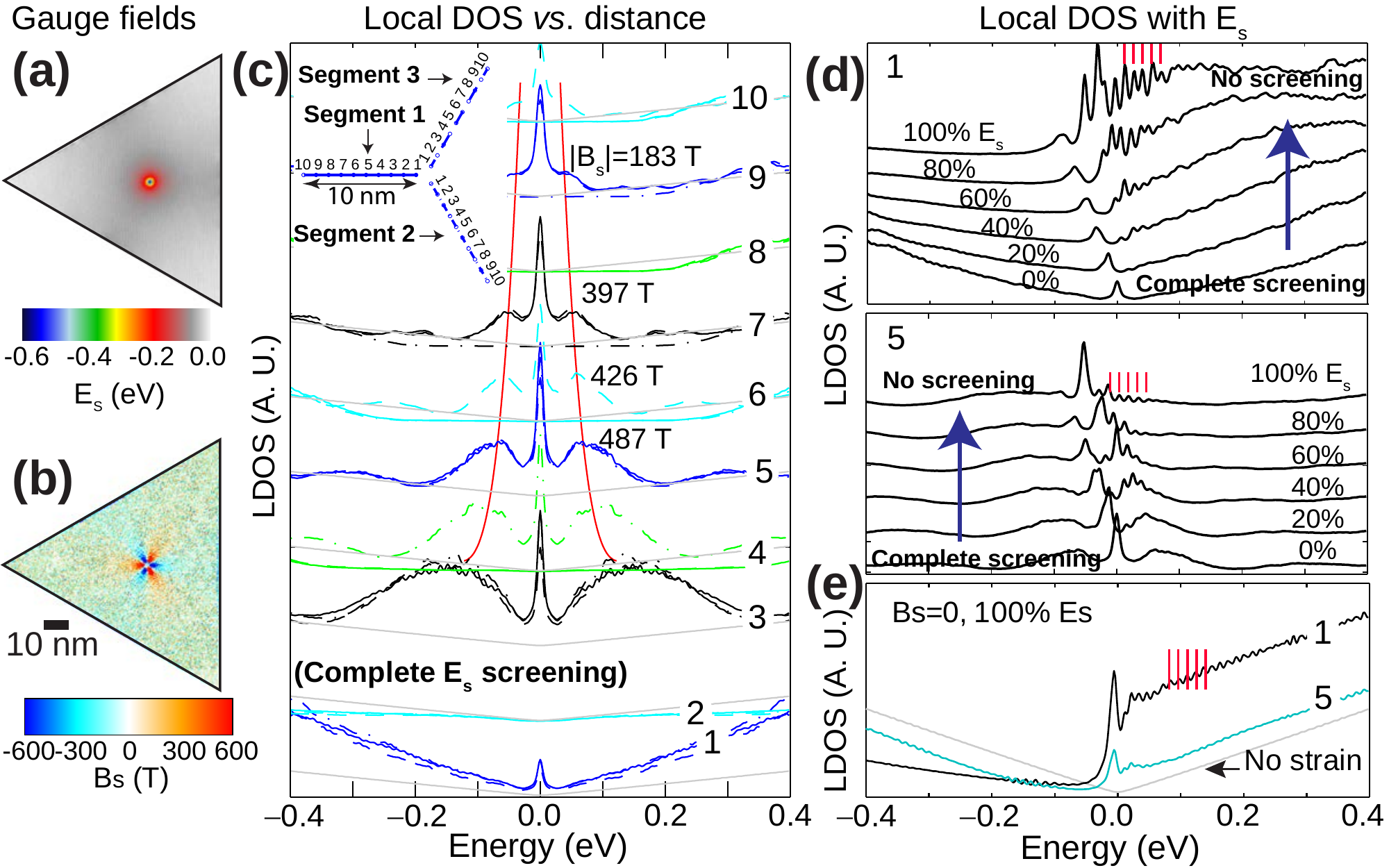}
 \caption{Color online. (a) $E_s$ and (b) $B_s$ ($\Gamma/l_1=35\%$). (c) Radial dependence of the LDOS ($E_s=0$). (d) Evolution of the LDOS at points 1 and 5 in (c) as $E_s$ is gradually turned on. (e) LDOS for $B_s=0$ and 100\% $E_s$.}\label{fig:F4}
\end{figure}

We display $\mathbf{B}_s(\mathbf{r})$ from Eqn~\eqref{eq:deltat} in Fig.~\ref{fig:F4}(b). $\mathbf{B}_s$ has the periodic angular dependency expected for a spherical extruder \cite{Ando2002,GuineaNatPhys2010,stroscio}. The ``pixelated'' texture of $B_s$ reminds us that  $\mathbf{A}_s$ is discrete in the present formalism.

\subsection{Local density of states}
The tight-binding Hamiltonian $\mathcal{H}$ and the LDOS are meaningful regardless of the scale of the mechanical deformation. Here we display the LDOS with a 5 meV energy resolution on membranes with rims containing three million atoms. To avoid rescaling \cite{GuineaNatPhys2010} we employed the Lanczos tight-binding method \cite{Zhengfei}.

We plot in Fig.~\ref{fig:F4}(c) the LDOS with $E_s$ turned off at ten radial positions (see inset in Fig.~\ref{fig:F4}(c)). For each radial position there are three curves, related by a 120$^{o}$ rotation. The curves are vertically offset for clarity. The gray v-shaped trendlines represent the DOS of unstrained graphene. We highlight a number of features: (i) A sharp zero LL, absent at some locations (points 2 and 8). (ii) Broad features in the LDOS, symmetric with respect to the zero level \cite{Crommie}; it is not clear if those correspond to a single LL or contain at least two broad LLs.
 From the energy locations for LLs $n=\pm 1$ and $n=0$ $B_s$ was estimated (assuming it uniform) and shown in some curves. Some curves are not symmetric under rotation (points 4, 6, 7, 9, and 10; not all three curves overlap). At locations 2 and 8 only a change in slope \cite{YWSon,deJuanPRL2012} is seen. Only when the pseudo-magnetic field is uniform should one expect the LLs to be sharp and position-independent.

\subsubsection{Relevance of the deformation potential in computing LDOS curves}
 In Fig.~\ref{fig:F4}(d) we gradually turn $E_s$ on (Fig.~\ref{fig:F4}(a)) at points 1 and 5. Importantly, equally-spaced peaks appear already at a screened $0.4E_s$, much like the equally-spaced peaks seen in Ref.~\cite{stroscio} at zero magnetic field. $E_s$ is not negligible in our system as it creates a confining well.

  To complete the study we plot in Fig.~\ref{fig:F4}(e) the LDOS at points 1 and 5, now setting $B_s=0$ (by using a membrane with no strain) and using $E_s$ from Fig.~\ref{fig:F4}(a): The plots in Fig.~\ref{fig:F4}(d) can not be understood as a simple superposition of plots in Figs.~\ref{fig:F4}(c) and those in Fig.~\ref{fig:F4}(d): Both $\mathbf{A}_s$ and $E_s$ are needed in computing the correct LDOS. Given the existence of dI/dV data obtained with local probes \cite{stroscio}, it is important for theoretical works to report LDOS data, complementing their reported $B_s$.

\section{Conclusions}
We have provided a theory for strain engineering valid beyond continuum elasticity, and strictly applicable for negligible curvature. We provide a measure to determine the extent to which mechanical distortions sublattice symmetry, in terms of changes in angles $\Delta \alpha$ and lengths $\Delta L$. For this we re-express the theory beyond continuum elasticity and explicitly on the atomic lattice.
Using this formalism, we studied triangular rippled graphene membranes under mechanical load by a sharp tip. Gauge fields were computed from atomic displacements alone.  We have found that rippled membranes will initially accommodate the extruder without increasing bond distances (graphene is a shell); neglecting this fact results in overestimated gauge fields. We also demonstrated in a simple way why no $K-$point dependent fields exist to first order. We studied the LDOS at many spatial locations. The scalar deformation potential $E_s$ gives rise to a number of equally-spaced peaks on the LDOS, even when partially screened.

\acknowledgments
We acknowledge computer support from HPC at Arkansas (\emph{RazorII}), and XSEDE (TG-PHY090002, \emph{Blacklight}, and \emph{Stampede}), and exchanges with B. Uchoa,  M. Vanevi\'c, L. Bellaiche, and M.~A. Kuroda.


\end{document}